\documentclass[final,5p,times,twocolumn,authoryear]{elsarticle}

\usepackage[authoryear]{natbib}
\usepackage{graphicx,amsmath,amssymb,amstext}
\usepackage{amsbsy,amsfonts,amsthm,color}
\usepackage[colorlinks,linkcolor=blue,citecolor=blue,urlcolor=blue ]{hyperref}
\usepackage[utf8]{inputenc}
\usepackage{float}
\usepackage{lipsum}
\usepackage{verbatim}
\usepackage[normalem]{ulem}
\usepackage{orcidlink}
\usepackage{soul}
\usepackage{subcaption}
\usepackage{pdflscape}

\newcommand{\aap}{A\&A}
\newcommand{\apj}{ApJ}
\newcommand{\apjl}{ApJL}
\newcommand{\mnras}{MNRAS}
\newcommand{\pasj}{PASJ}
\newcommand{\ssr}{Space Sci. Rev.}
\newcommand{\nar}{New Astron. Rev.}
\newcommand{\aapr}{A\&A Rev.}

\newcommand{\apss}{Ap\&SS}

\journal{High Energy Astrophysics}

\begin{document}

\begin{frontmatter}


\title{Constraints on cyclotron features and accretion regime in the \\ high-mass X-ray binary 4U 1700--37 from {\it NuSTAR}}

\author[first,second]{L. West-Ocampo\,\orcidlink{0009-0008-5828-7397}}
\ead{lwest@fcaglp.unlp.edu.ar}

\author[second]{F. A. Fogantini\,\orcidlink{0000-0001-5628-9120}}

\author[third,fourth]{E. A. Saavedra\,\orcidlink{0000-0003-4841-8302}}

\author[first,second,fifth]{J. A. Combi\,\orcidlink{0000-0002-2565-5025}}

\author[first,second]{F. García\,\orcidlink{0000-0001-9072-4069}}

\author[sixth]{P. L. Luque-Escamilla\,\orcidlink{0000-0002-3306-9456}}

\author[fifth]{J. Martí\,\orcidlink{0000-0001-5302-0660}}

\author[seventh]{S. Chaty\,\orcidlink{0000-0002-5769-8601}}

\author[eighth]{J. F. Albacete-Colombo\,\orcidlink{0000-0001-8398-0515}}

\affiliation[first]{
  organization={Facultad de Ciencias Astronómicas y Geofísicas, Universidad Nacional de La Plata},
  addressline={B1900FWA La Plata},
  city={La Plata},
  country={Argentina}
}

\affiliation[second]{
  organization={Instituto Argentino de Radioastronomía (CCT La Plata, CONICET; CICPBA; UNLP)},
  addressline={C.C. 5},
  city={Villa Elisa},
  postcode={1894},
  country={Argentina}
}

\affiliation[third]{
  organization={Instituto de Astrofísica de Canarias (IAC)},
  addressline={Vía Láctea s/n},
  city={La Laguna},
  postcode={38205},
  state={S/C de Tenerife},
  country={Spain}
}

\affiliation[fourth]{
  organization={Departamento de Astrofísica, Universidad de La Laguna},
  city={La Laguna},
  postcode={38205},
  state={S/C de Tenerife},
  country={Spain}
}

\affiliation[fifth]{
  organization={Departamento de Física (EPS), Universidad de Jaén},
  addressline={Campus Las Lagunillas},
  city={Jaén},
  postcode={23071},
  country={Spain}
}

\affiliation[sixth]{
  organization={Departamento de Ingeniería Mecánica y Minera (EPSJ), Universidad de Jaén},
  addressline={Campus Las Lagunillas},
  city={Jaén},
  postcode={23071},
  country={Spain}
}

\affiliation[seventh]{
  organization={Université Paris Cité, CNRS, Astroparticule et Cosmologie},
  city={Paris},
  postcode={75013},
  country={France}
}

\affiliation[eighth]{
  organization={Universidad de Río Negro, Sede Atlántica - CONICET},
  city={Viedma},
  postcode={CP8500},
  state={Río Negro},
  country={Argentina}
}

\cortext[cor1]{Corresponding author: Lautaro West-Ocampo}


\begin{abstract}
{4U~1700--37 is a wind-fed high-mass X-ray binary comprising a compact object, most likely a neutron star, accreting from the O6.5\,Iaf$^{+}$ supergiant HD~153919. Coherent pulsations have not been firmly detected and the magnetic field strength remains uncertain, with previous {\it NuSTAR} studies reporting only marginal candidate cyclotron resonant scattering features (CRSFs).}
{We analyses all available {\it NuSTAR} observations of 4U~1700--37 to characterize its hard X-ray timing and broadband spectral properties, test the robustness of candidate CRSFs against different continuum models and epochs, and constrain the magnetic field and accretion regime of the compact object.}
{We perform a homogeneous timing and spectral analysis of two {\it NuSTAR} observations, modeling time-averaged and intensity-resolved spectra with several continua commonly used for accreting pulsars, and use spectral simulations to quantify the significance and model dependence of putative CRSFs.}
{No coherent pulsations are detected and we constrain the pulsed fraction to be <1.5$\%$.
The spectra are well described by an absorbed blackbody plus cut-off power-law continuum, but show shallow absorption-like residuals around ${\sim}20$~keV and at higher energies (${\sim}40$–$50$~keV). These features improve the fits and can reach high formal significances for some continua, yet do not constitute firm, model-independent CRSF detections in our baseline analysis. Intensity-resolved spectroscopy hints at shifts of the apparent line centroid with flux.}
{Interpreted as cyclotron features, these results favour a neutron-star magnetic field $B\sim(1.7$–$4.4)\times10^{12}$~G and a quasi-spherical subsonic accretion regime with an equilibrium spin period $P_{\rm eq}\sim1.9$~ks. Even without a secure CRSF detection, the homogeneous multi-epoch analysis provides quantitative constraints on the magnetic field and accretion physics of 4U~1700--37 and helps reconcile previously discrepant line-energy measurements.}
\end{abstract}



\end{frontmatter}


\section{Introduction} \label{sec:intro}

A number of significant magnetized neutron stars (NSs) in binary systems exhibit distinctive spectral features in their high-energy X-ray spectra \citep{Chaty2022}.
These features, known as cyclotron lines or cyclotron resonant scattering features (CRSFs), primarily appear as absorption features in the 10 to 100 keV energy range \citep{Staubert2019}.
Cyclotron lines are produced near the magnetic poles of accreting NSs, where the energetic plasma from infalling material converts kinetic energy into heat and radiation. Within this hot and highly magnetized environment, electrons assume discrete energies known as Landau levels, which are determined by their motion perpendicular to the magnetic field. When photons interact with these electrons through resonant scattering, features in the X-ray spectrum are observed as absorption lines \citep{Lutovinov2017}. 
The detection of such features allows for the measurement of the magnetic field strength through the centroid line energies $E_{\rm cyc}$ \citep{Staubert2019}.

4U~1700--37 (hereafter 4U1700) is a transient, eclipsing high‑mass X‑ray binary (HMXB), first detected by the \textit{Uhuru} satellite in 1970 \citep{Jones1973ApJ...184L..65J}. 
Its optical counterpart, the runaway O6.5\,Iaf$^{+}$ supergiant star HD~153919, was identified shortly thereafter \citep{Hutchings1974ApJ...192..677H}, later allowing to classify the system as a supergiant X‑ray binary (sgHMXB) \citep{Chaty2008,Coleiro2013}. 
Gaia~DR3 places the system at a distance of $\sim$1.9~kpc, likely within the Sco~OB1 association \citep{GaiaDR3Summary}. The source ranks among the brightest persistent Galactic hard X-ray emitters monitored by \textit{INTEGRAL} and \textit{Swift}/BAT \citep{Walter2015A&ARv..23....2W}.
The binary completes an eccentric orbit ($e\simeq0.22$) every 3.412~d, with an inclination of ${\sim}66^\circ$ \citep{Rubin1996}, producing deep X‑ray eclipses and modulations clearly visible in the X-ray light curves \citep{Islam2016MNRAS.461..816I}. 

The compact object in the system is thought to be most likely a NS \citep{Clark2002, vandermeer2005}.
In earlier studies, \cite{1978Branduardi} and \cite{1978Matilsky} proposed pulsations with periods of 24 and 97 minutes, respectively, based on observations from Copernicus and SAS-3. However, \cite{1979Hammerschlag} suggested that the 97-minute period might be an instrumental artefact. Later, \cite{1984Murakami} detected a possible 67.4~s pulsation during the brightest flare observed with {\it Tenma} \citep{1984Simpson}. None of these periods have been confirmed by subsequent studies \cite[e.g.,][]{Gottwald1986MNRAS.222P..21G}.
Although pulsations are expected in this type of HMXBs, some systems do not exhibit them \citep[e.g.,][]{2020Sguera, 2018Masetti}.

\textit{Chandra} spectroscopy during a major flare revealed Compton cooling consistent with a NS accretor of mass $M_{\mathrm X}\simeq2.4\,M_\odot$ \citep{MartinezChicharro2018MNRAS.473L..74M}, and broadband spectral analysis using \textit{NuSTAR} and \textit{AstroSat} \cite{Agrawal2006A} observations showed marginal detections of CRSFs at ${\sim}16$~keV \citep{Bala2020MNRAS.493.3045B}.  

Despite these efforts, the existence, energy and stability of any CRSFs in 4U1700 remain debated, and the implications for the NS magnetic field and accretion regime have not been quantified in a homogeneous way. In particular, previous hard X-ray studies were based on single epochs, single-continuum assumptions and earlier calibration releases, which makes it difficult to assess the robustness of the reported features and to compare different campaigns on equal footing.

In this work we present a homogeneous temporal and spectral analysis of all available \textit{NuSTAR} observations of 4U1700. In Section~\ref{sec:data} we describe the \textit{Swift}/BAT and \textit{NuSTAR} datasets, the updated calibration, and the data reduction strategy, including the treatment of instrumental edges. In Section~\ref{sec:results} we present the timing analysis, continuum modelling, and a systematic search for CRSFs in both epochs, together with intensity-resolved spectroscopy. In Section~\ref{sec:discussion} we interpret our findings in the context of quasi-spherical subsonic accretion onto a magnetized NS and derive constraints on the magnetic field and equilibrium spin period. Finally, in Section~\ref{sec:conclusions} we summarize our main results and discuss the observational requirements to confirm or refute the putative cyclotron features.


\section{Observation and Data Analysis} \label{sec:data}

\begin{figure}
    \centering
    \includegraphics[width=0.475\textwidth]{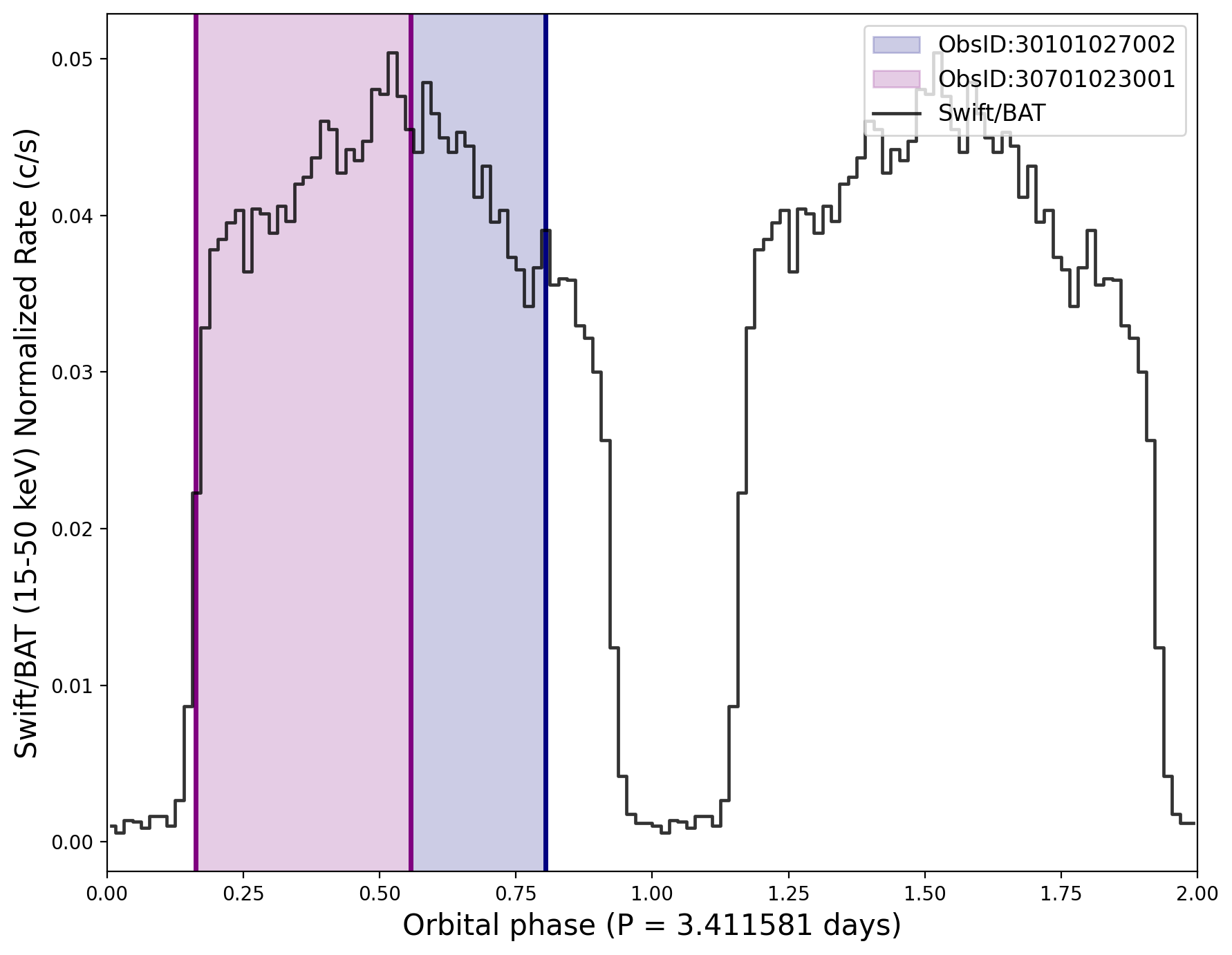}
    \caption{Swift/BAT folded light curve of 4U1700 using 64 orbital phase bins. The data were folded using a orbital period of $3.411581$~d \citep{Falanga2015A&A...577A.130F} and a reference epoch $T_0 [\rm MJD]= 57446.55$ \citep{Bala2020MNRAS.493.3045B}. The \textit{NuSTAR}\ observations of 4U1700\ used in this work are represented by the purple and blue colored stripes, spanning $\sim$39\% and $\sim$25\% of the orbital period, respectively.}
    \label{fig:bat}
\end{figure}

\subsection{Swift/BAT data}

The Burst Alert Telescope (BAT; \citealt{Barthelmy2005}) is a wide-field, hard X-ray telescope designed to detect and locate gamma-ray bursts (GRBs), which is part of the Neil Gehrels Swift Observatory \citep{Krimm2004}, a multi-wavelength observatory launched by NASA in 2004. The BAT uses a coded aperture mask to achieve a large field of view and high sensitivity in the 15-150 keV energy range. 

In this paper we consider the full orbital light curve of 4U1700 available up to December 1st, 2024 in the Swift/BAT service\footnote{\url{https://swift.gsfc.nasa.gov/results/transients/4U1700-377/}}, a public website where more than 1000 light curves of hard X-ray sources are available, spanning close to 20 years of continuous data. 

On \hyperref[fig:bat]{Figure~\ref{fig:bat}} we show the folded orbital Swift/BAT lightcurve of 4U1700. We used the orbital period of 3.411581~d \citep{Falanga2015A&A...577A.130F} and the reference epoch $T_0 [\rm MJD]= 57446.55$ \citep{Bala2020MNRAS.493.3045B}. We also show with a vertical band the {\it NuSTAR} observation analyzed in this work, which spans almost 40\% of the orbital period after the eclipse.

\subsection{\textit{NuSTAR} data}

The {\sl Nuclear Spectroscopic Telescope Array} (\textit{NuSTAR}) is a space-based X-ray observatory consisting of two focal plane modules, FPMA and FPMB, equipped with 2$\times$2 solid-state CdZnTe detectors each and operating in the 3--79~keV energy range \citep{Harrison2013}.

\textit{NuSTAR} observed 4U1700 on March 1st 2016 for ${\sim}38$~ks (ObsID: 30101027002) and again on January 27th 2022 for ${\sim}50$~ks (ObsID: 30701023001). We processed both datasets in a homogeneous way using the \textit{NuSTAR} Data Analysis Software ({\textsc NuSTARDAS}) distributed within the {\textsc HEASOFT}~v6.34 package. The event files were produced with {\textsc nupipeline} and the most recent calibration database (CALDB v.20240229), and barycentric corrections were applied using the corresponding clock correction file.

In order to filter out events from the South Atlantic Anomaly (SAA) passages, we looked at the individual observation report \footnote{\url{https://nustarsoc.caltech.edu/NuSTAR_Public/NuSTAROperationSite/SAA_Filtering/SAA_Filter.php}}. 
We extracted cleaned event files using {\textsc saacalc=3}, {\textsc saamode=STRICT} and {\textsc tentacle=YES} SAA parameters. 

Source and background spectra, light curves and response files were extracted for FPMA and FPMB using {\textsc nuproducts}, adopting circular source regions of 60~arcsec centred on 4U1700 and background regions of the same size on nearby, source-free detector areas. The same extraction regions and filtering criteria were used for both epochs to ensure a consistent comparison. Background subtraction of each camera module and later addition of the corrected light curves were performed using the {\textsc LCMATH} task.
We generated source and background spectra and the ancillary and response matrices with \textsc{nuproducts}, using the same regions as for the light-curve extraction. To properly use the convolution models, we extended the energy array using \textrm{XSPEC}'s \textsc{energies} command.

The \textit{NuSTAR} spectrum of 4U1700 (obsID:30701023001) was affected by sharp instrumental absorption features (Zn and Cd K-edges) present in the FPM detectors that could not be removed using the standard pipeline tools (\citealt{Madsen2015}; see also \citealp{ Iaria2024A&A...683A..79I, Malacaria2025A&A...699A.288M, Parra2025arXiv250400991P}). To account for this, we included two absorption edges, modelled by the \textsc{edge} component in \textrm{XSPEC}, with energies fixed at approximately ${\sim}9.3$~keV and ${\sim}26.9$~keV, corresponding to the K-edges of Zn and Cd, respectively. The optical depths of these absorption edges are of the order of 0.01$-$0.02 in both cases.


\section{Results} \label{sec:results}

The results presented in the following subsections are based on a detailed spectral and temporal analysis of the \textit{NuSTAR} observation with obsID 30701023001. For comparison, we also analysed the archival observation (obsID: 30101027002) to provide additional context for our findings.

\subsection{Temporal analysis}

\begin{figure}[ht]
    \centering
    \includegraphics[width=\columnwidth]{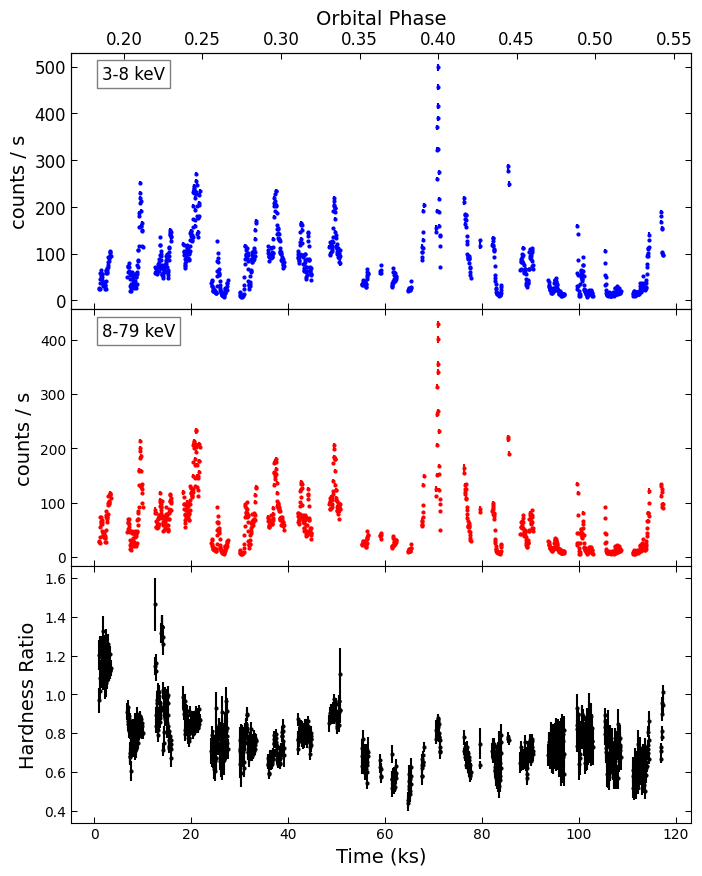}
    \caption{\textit{NuSTAR} background-subtracted light curves of 4U1700 extracted with a bin time of $50~\mathrm{s}$.
    Top panel: 3--8~$\mathrm{keV}$ light curve.
    Middle panel: 8--79~$\mathrm{keV}$ light curve.
    Bottom panel: hardness ratio.}
\label{4U1700_lc}
\end{figure}

The 3--70 keV background-corrected light curve exhibits flaring activity on timescales of hundreds to thousands of seconds, with count rates increasing by up to a factor of three in the 50~s binned light curves. We defined two energy bands, a soft band (3--8 keV) and a hard band (8--79 keV), defined in a way to have a comparable number of total counts per band. 
Figure~\ref{4U1700_lc} shows both soft and hard X-ray light curves (top and middle panels), as well as the hardness ratio light curve (bottom panel), defined as the ratio between hard and soft X-ray lightcurves.
Figure~\ref{4U1700_lc} also reveals a major X-ray flare lasting ${\sim}$8~ks and shorter flares of ${\sim}$1~ks, with count rates increasing by a factor of ${\sim}$3 relative to the persistent emission, that has an average count rate of ${\sim}$53~counts\,s$^{-1}$ in the full spectrum lightcurve. 
Throughout the observation, the hardness ratio remains relatively stable with an average value of $0.8\pm0.1$, despite the intensity variations in each energy band. The extracted light curves indicate that the hard emission predominates during the initial 5~ks, while the soft band becomes dominant for the remainder of the exposure, exhibiting a ${\sim}$40~ks long modulation.

To explore the variability of 4U1700 in a more detailed manner, we extracted {\it NuSTAR} power spectra density (PDS) in the 3--70~keV energy range from the cleaned event files using the {\sc powspec} task from {\sc HEASoft} package. We used a time resolution of 0.05~s, which sets the Nyquist frequency to 40~Hz. We split the data into 409.6~s segments, yielding a minimum frequency of ${\sim}2.4\times 10^{-3}$~Hz. The power spectrum is dominated by broadband powerlaw-like noise up to ${\sim}$0.1~Hz, frequency at which the Poisson noise starts to dominate. No significant features are seen above noise.
We also computed the Lomb-Scargle periodogram \citep{Lomb1976, Scargle1982} to search for periodicities in the 0.01--5000~s range, relevant for the slow equilibrium periods expected in wind-fed settlers. No significant candidates were detected.

To further investigate the presence of coherent signals, we utilized the \texttt{HENDRICS} package \citep{hendrics}. We first performed a blind search for pulsations using \texttt{HENaccelsearch} tool across the 0.03$-$10~Hz frequency band, allowing for frequency derivatives up to $10^{-9}$~Hz~s$^{-1}$.  
The lower frequency limit was set to $0.03$~Hz because the spectrum below this cutoff is dominated by strong red noise ($S(f) \propto f^{-\alpha}$), which raises the local detection threshold and masks potential weak periodic signals.

Since no significant candidates were detected, we estimated the sensitivity of our observation by calculating upper limits on the pulsed fraction (PF) using the \texttt{HENz2vspf} tool in the 3--70~keV band.
This method shuffles event times and simulates pulsations with random PFs, calculating the maximum $Z^2$ value within a narrow frequency interval around the injected signal. The process is repeated $N_{\rm trial}$ times, where PF is defined as the ratio of the pulsation amplitude to the mean flux. Using $10^4$ simulations, we derived 90\% confidence upper limits on the PF, yielding upper limits of approximately 1.36$\%$ and 1.47$\%$ for FPMA and FPMB, respectively.

\subsection{Spectral analysis}

We performed X-ray spectral analysis using \textrm{XSPEC} v12.14.1 \citep{Arnaud1996}, considering the 3--70~keV energy range. The time-averaged spectra were grouped using the optimal binning method of \cite{Kaastra2016} with a minimum of 30 counts per bin to properly use $\chi^2$ statistics.

The \textit{NuSTAR} time-averaged spectra of 4U1700 shows a typical heavily absorbed powerlaw-like continuum, with narrow features like the Fe K$\alpha$ emission line complex at $\sim$6.4~keV. To find the best continuum model, we fitted the FPMA+B spectra with different variations of thermalized continua and thermal plus non-thermal continuum models frequently used to describe this kind of systems \citep[e.g.,][]{Staubert2019, Sidoli2018, 2023Fornasini}. For the former, we used a Comptonized black-body spectrum (\textsc{thcomp*bbody}; \cite{Zdziarski2020}), which depends mainly on the blackbody seed photon spectrum temperature $kT_{bb}$, as well as on the electron scattering plasma temperature $kT_e$, the low-energy power-law index $\Gamma_\tau$, and the covering fraction or Comptonization fraction. For the non-thermal models, we tried four widely used phenomenological models for X-ray pulsars: \textsc{bbody+cutoffpl}, \textsc{bbody+powerlaw $\times$ highecut}, \textsc{bbody+nthcomp}, and \textsc{npex} \citep{Mihara1995}. The \textsc{npex} model consists of two power-law components, one with a positive and the other with a negative photon index, both affected by the same high-energy cutoff.

To model the Fe feature at 6.4~keV, we used a Gaussian component. To account for both Galactic and local absorption, we used the \textsc{pcfabs} model, with abundances of \cite{Anders1989} and cross section from \cite{Verner1996}. We used a calibration constant to account for cross-calibration uncertainties between both \textit{NuSTAR} instruments FPMA and FPMB. We checked that this constant remained within the typical 2--3\% cross-calibration uncertainty range usually observed between the modules \citep{Madsen2015}.

To estimate the parameter uncertainties, we performed Monte Carlo Markov Chain (MCMC) simulations. Using \textrm{XSPEC}'s {\sc chain} task, we produced chains of length $10^{6}$, with walkers equal to 16 times the number of free parameters. We checked that the trace plot of each parameter successfully converged, ensuring all walkers explored the parameter space sufficiently and uniformly (see \citealt{Fogantini2023, saavedragx13} for more details).

\begin{figure}[ht]
    \centering
    \includegraphics[width=\columnwidth]{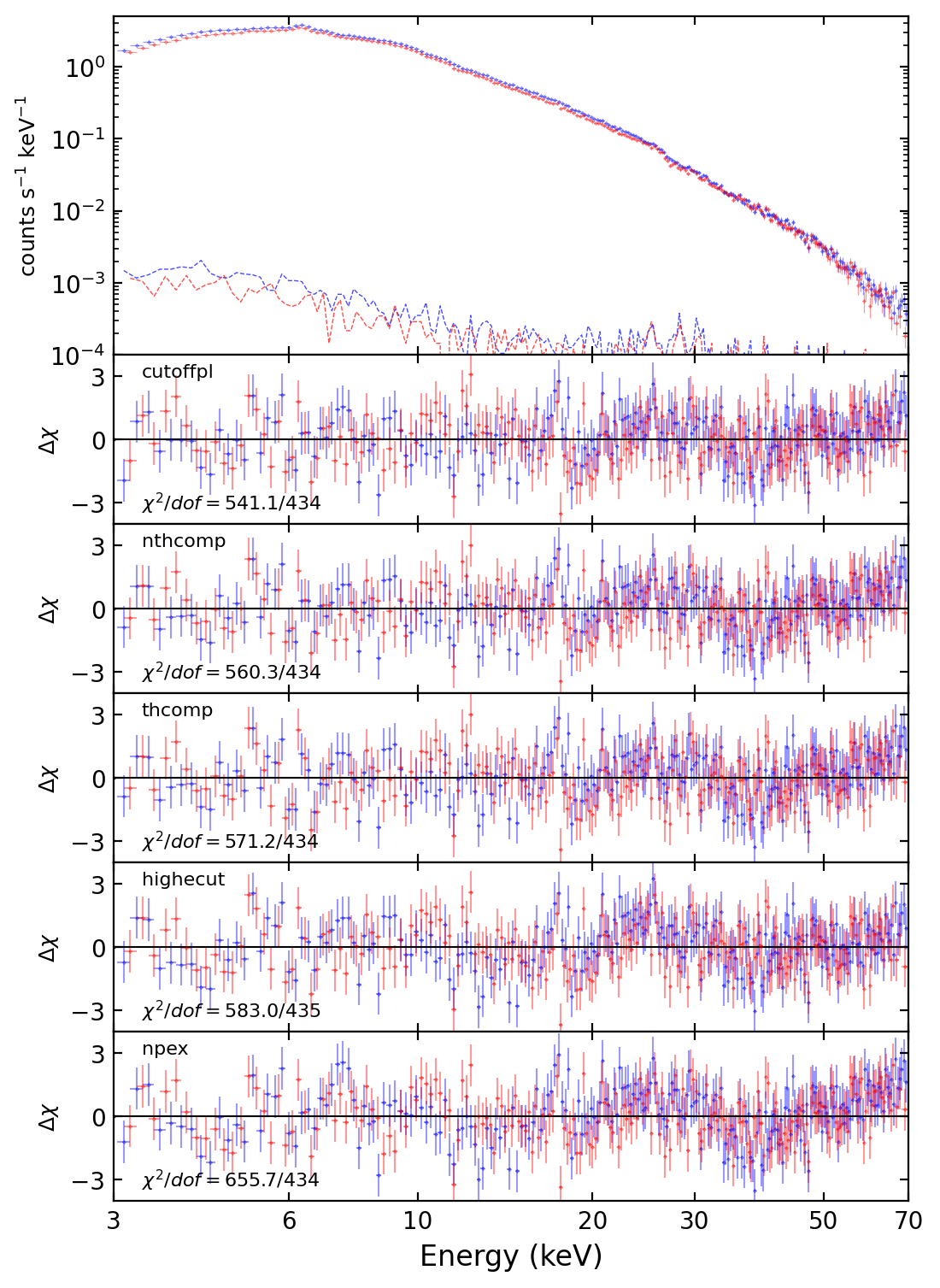}
    \caption{Top panel: FPMA/B background-subtracted spectra of 4U1700. Background spectra are also shown in dashed lines. Second-to-last panels: residuals obtained from fitting different continuum models. The corresponding model label is indicated in each panel. All models contain a Gaussian line component at ${\sim}$6.3~keV. The total $\chi^{2}$ and degrees of freedom are shown on the bottom left part of each panel. 
    A modulation in the residuals above 30~keV can be seen for all the tested models.}
\label{4U1700_spec}
\end{figure}

We show in Fig.~\ref{4U1700_spec} the best-fitting model residuals from the five tested continuum models applied to the time-averaged \textit{NuSTAR} spectra. In terms of statistical values, we find that the \textsc{bbody+cutoffpl} model provides the best fit, with $\chi ^2$/dof = 541.1/434${\sim}$1.25.  In Table~$\ref{parameters}$, we present the best-fitting parameters and uncertainties of the tested continuum models.

From Table~\ref{parameters} we see that the absorption model {\sc pcfabs} yields a hydrogen column density of $N_{\rm H}{\sim}18\times 10^{22}~\mathrm{cm}^{-2}$, indicating significant absorption likely caused by local absorbing material (e.g., wind clumps) around the X-ray source, with a covering factor of $f_{\rm cov}{\sim}0.44$. 
The black body temperature is of ${\sim}1.64~\mathrm{keV}$, consistent with thermal emission from the NS surface. Using the normalisation $K_{\rm{BB}}$ and the standard relation for the blackbody flux, we estimated the radius of the emitting region using a distance of $\sim$1.9~kpc, yielding a radius of $\sim$400~m, suggestive of emission from the polar caps or a hotspot on the NS \citep[see eg.,][]{Jaisawal2018MNRAS, PaloMere2007}. The blackbody component contributes an unabsorbed flux of $3.1\pm0.2\times10^{-10}~\mathrm{erg\,cm^{-2}\,s^{-1}}$ in the 3--70~keV energy band, representing ${\sim}8\%$ of the total emission.
The {\sc cutoffpl} component introduces a non-thermal contribution, with a photon index ($\Gamma{\sim}1.05$) that models the high-energy part of the spectrum, characteristic of high-energy processes like inverse Compton scattering in a corona of energetic electrons.
The power-law component yields an unabsorbed flux of $3.59_{-0.04}^{+0.03} \times 10^{-9}~\mathrm{erg\,cm^{-2}\,s^{-1}}$ in the 3--70~keV band, contributing ${\sim}92\%$ of the total emission at higher energies.
And the cut-off energy of $E_{\rm cut}{\sim}17.1~\mathrm{keV}$ indicates the energy scale at which the non-thermal emission, likely driven by Comptonization in the accretion column or corona, begins to decline, consistent with the high-energy spectral properties of accreting NSs.
And the $\textsc{gaussian}$ component models a Fe K$\alpha$ emission line at ${\sim}6.4~\mathrm{keV}$ with a width of ${\sim}0.02~\mathrm{keV}$, consistent with fluorescent iron emission from the accreted material.

As shown in Fig.~\ref{4U1700_spec}, however, all continua leave broad, absorption-like residuals above ${\sim}20$~keV, extending to the high-energy end of the spectrum. This motivates a dedicated search for candidate cyclotron resonance scattering features on top of the different continua and underlines the need to quantify how sensitive any putative line is to the choice of underlying continuum model.

\begin{table*}[!t]
\fontsize{11}{3}
\renewcommand{\arraystretch}{2.0}
\centering
\begin{tabular}{p{3.5cm}p{2.5cm}p{2.5cm}p{2.5cm}p{2.5cm}p{2.5cm}}
\hline
Parameter & CutoffPl & NThComp  & ThComp  & HighECut & NPEX \\
\hline \hline
$N_{\rm H}~\rm{[10^{22}~cm^{-2}]}$            & $18 \pm 1$ & $35 \pm 2$ & $38 \pm 2$ & $13\pm1$ & $17 \pm 1$ \\
$f_{\rm cov}$                            & $0.44 \pm 0.02$ & $0.31 \pm 0.01$ & $0.3 \pm 0.01$ & $0.54 \pm 0.01$ & $0.51 \pm 0.01$ \\ \hline
$kT_{BB}$~[keV]                         & $1.64_{-0.02}^{+0.03}$ & $1.19_{-0.02}^{+0.03}$ & $1.22 \pm 0.02$ & - & - \\
$K_{BB}$~[$10^{-4}$]                    & $43 \pm 3$ & $3_{-2}^{+4}$ & $232_{-3}^{+4}$ & -&  - \\ \hline
$\Gamma^{+}$                            & $1.05 \pm 0.03$ & - &   - & $1.33 \pm 0.01$ & $0.84 \pm 0.01$ \\
$\Gamma^{-}$                            & - & - &  - & - & $-2.0_{-0.1}^{+0.2}$ \\
$E_{\rm cut}$~[keV]                      & $17.1 \pm 0.4$ & - &   - & $6.4 \pm 0.1$ & $8.4 \pm 0.1$ \\
$\text{Norm}^{+}$                        & $0.18 \pm 0.01$  & $0.041_{-0.002}^{+0.001}$ &   - & $0.25 \pm 0.01$ & $0.22 \pm 0.01$ \\
$\text{Norm}^{-}$                        & - & - & -& - & $(21 \pm 1){\times}10^{-6}$ \\
$F~\rm{[10^{-9}~erg \, cm^{-2} \, s^{-1}]}$     & $3.91_{-0.02}^{+0.01}$ & $3.92\pm0.02$ & $3.93_{-0.02}^{+0.01}$ & $3.87_{-0.01}^{+0.02}$ & $3.97_{-0.02}^{+0.01}$ \\ \hline
$E_{fold}$~[keV]                         & - & - &  - & $20.3 \pm 0.3$ & -   \\ \hline
$\Gamma_\tau$                            & - & $1.94\pm0.01$ & $1.94 \pm 0.01$ &  - & - \\
$kT_{e}$~[keV]                           & - & $10.5 \pm 0.1$ & $10.2 \pm 0.1$ & -&  - \\
$f_{\rm comp}$                           & - & - & $0.99_{-0.02}^{+0.01}$ & - & - \\ \hline
$E_{\rm K\alpha}$~[keV]                  & \multicolumn{5}{c}{$6.32\pm0.01$} \\
$\sigma_{\rm K\alpha}$~[keV]             & \multicolumn{5}{c}{$0.02\pm0.01$}  \\
$\text{Norm}~[10^{-4}]$                  & $12.4 \pm 0.4$ & $12.5_{-0.5}^{+0.4}$ & $12.5_{-0.5}^{+0.4}$ & $11.5 \pm 0.4$ & $12.8 \pm 0.4$ \\ \hline
$\text{Constant}$                        & \multicolumn{5}{c}{$1.026 \pm 0.001$}  \\
$\chi^2/{\rm dof}$                       & $541.1/434$ & $560.3/434$ & $571.1/434$ & $583/435$ & $656.7/434$ \\
\hline \hline
\end{tabular}
\caption{Best fit parameters obtained from fitting different continuum models to the time-averaged \textit{NuSTAR} spectra of 4U1700. Reported uncertainties correspond to a significance level of $1\sigma$ derived from MCMC chains. Norm$^{\pm}$ are in units of [photons\,keV$^{-1}$\,cm$^{-2}$\,s$^{-1}$]. The positive/negative supra-indexes in Norm and $\Gamma$ refer to the positive/negative power-law component in the \texttt{npex} model. In the remaining models, only the positive supra-index parameters should be regarded as relevant.} 
\label{parameters}
\end{table*}

\begin{figure}[ht]
    \centering
    \includegraphics[width=\columnwidth]{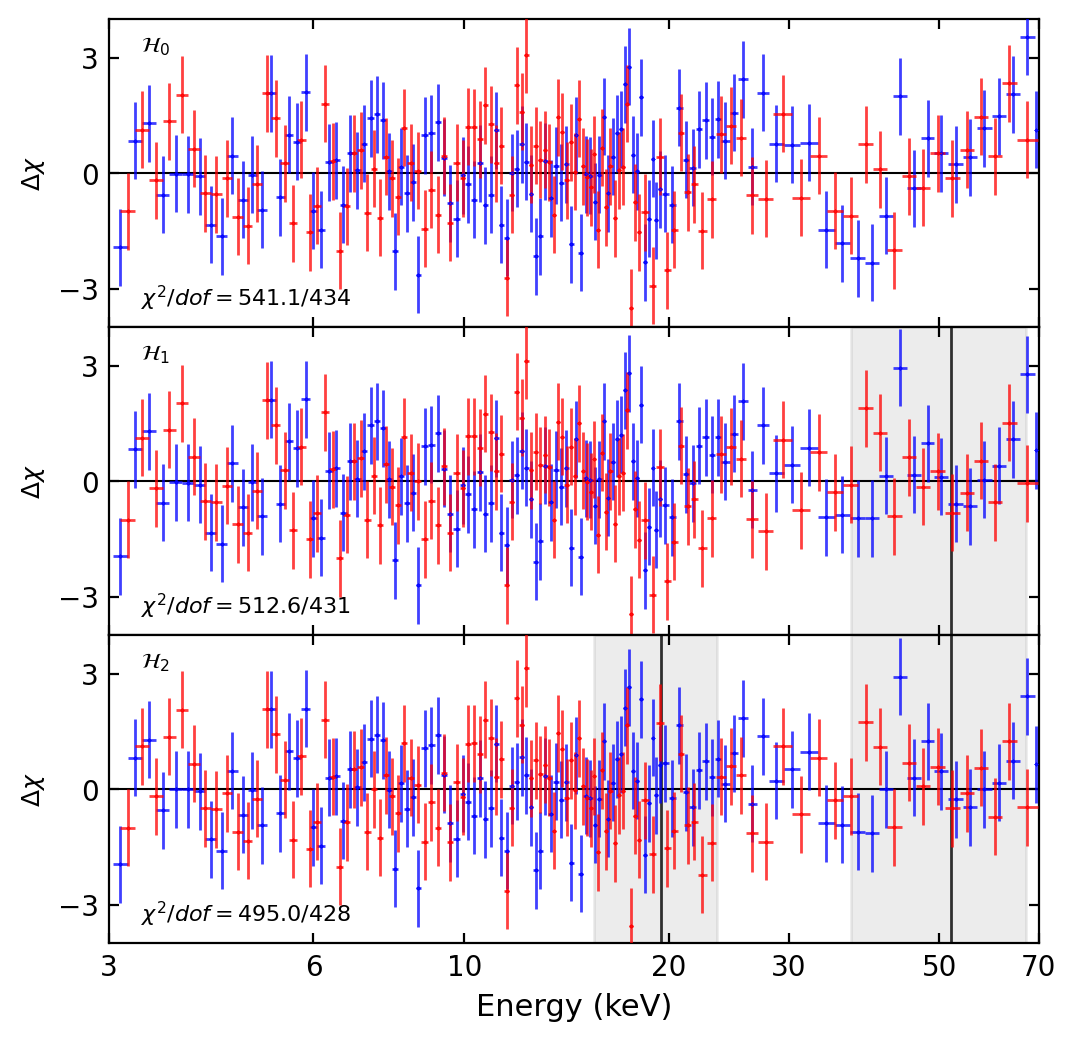}
    \caption{\textit{NuSTAR} residuals of 4U1700's using the {\sc cutoffpl} model as baseline. The top panel shows only the residuals of the continuum and Fe line contributions ($\mathcal{H}_0$). The middle panel shows the residuals of the continuum affected by a single \textsc{gabs} component ($\mathcal{H}_1$) at ${\sim}50$~keV. The bottom panel also includes a second \textsc{gabs} component ($\mathcal{H}_2$) at ${\sim}20$~keV. Total $\chi^{2}$ and dof are indicated on the bottom left part of each panel. The vertical gray stripes indicate the CSRFs candidates with their corresponding centroids and widths. The data were re-binned for better visualization.}
\label{4U1700_res}
\end{figure}

\subsubsection{Testing for CRSFs}

We searched for CRSFs in the \textit{NuSTAR} spectra of 4U1700. To model CRSFs, we used the \textsc{gabs} model in \textrm{XSPEC}, which depends on three parameters: the line centroid energy $E_{\rm cyc}$, the line width $\sigma_{\rm cyc}$, and the strength, parameter which can be used to derive the optical depth at the centroid. We performed the same analysis using the $\textsc{cyclabs}$ model, and it yielded results consistent with those obtained with $\textsc{gabs}$. Therefore, we will only report the results obtained with {\sc gabs}.

Using the \textrm{XSPEC} \textsc{steppar} command, we explored the $\chi^2$ space with line centroids in the 10–-70~keV range in steps of 0.5~keV, line widths between 0.5--5~keV with 0.1~keV steps, and strength of 1~keV. We found two CRSFs candidates in the {\it NuSTAR} spectra: the first and most significant at ${\sim}$52~keV, and a second at ${\sim}$19.5~keV. Performing a F-test to compare the \textsc{cutoffpl} and \textsc{gabs*cutoffpl} models, we obtained p-values of $3.4\times10^{-5}$ for the 52~keV feature and $0.002$ for the 19.5~keV feature.

To assess the significance of these features, we performed over $10^5$ spectral simulations using the \textrm{XSPEC} \textsc{simftest} command. We compared the F-values from simulated spectra to the F-value from the real data to determine the minimum detection significance. Following \citet{Hurkett2008}, we calculated the F-value as $F = (\nu_{\rm 0} / \Delta\nu) \times (\Delta\chi^2 / \chi^2_{\rm 1})$, where subindexes $\{0,1\}$ denote the null hypothesis ($\mathcal{H}_0$, {\sc cutoffpl} continuum only) and the tested hypothesis ($\mathcal{H}_1$, continuum affected by a single \textsc{gabs} component), with total $\chi^2$ and $\nu$ degrees of freedom, respectively. The significance level is derived from the $p$-value, defined as the fraction of simulated spectra with F-values exceeding that of the real data. The significance is expressed as $(1 - p)$ in per-cent units.

In Fig.~\ref{4U1700_res} we present the residuals of the best fitting tested continuum, in which the top panel shows the null hypothesis $\mathcal{H}_0$ and the middle panel shows residuals from $\mathcal{H}_1$ (which corresponds to the most significant candidate, at ${\sim}$52~keV). To test the presence of a second CRSF component ($\mathcal{H}_2$) we proceeded to repeat the same algorithm as stated before, but using $\mathrm{H_1}$ as null hypothesis. The residuals of $\mathcal{H}_2$ are shown on the bottom panel. In Table~$\ref{cyctable}$ we present the detection levels of both \textsc{gabs} components, with their best-fitting parameters and uncertainties. Both scenarios yield significance levels below $3\sigma$.

After investigating the presence of CRSFs with our best-fitting continuum model, we derived lower limits on their detection significance (Table~\ref{cyctable}). The sequential fitting procedure first adds a line at $E_{\mathcal{H}1}\simeq52\ \mathrm{keV}$, improving the fit with a probability of $\sim95\%$. Introducing a second feature at $E_{\mathcal{H}2}\simeq19.5\ \mathrm{keV}$ raises the probability only marginally to $\sim95.5\%$. In both cases the significance remains well below the $3\sigma$ threshold, so neither feature can be considered a firm detection. However, alternative continuum models that have higher reduced $\chi ^2$ statistics (see Table~\ref{parameters}) yield higher detection significances for these features in the spectrum of 4U1700. Performing the same analysis with other continuum models, we obtained significances of $99.9\%$ and $98.42\%$ for $\mathcal{H}_1$ and $\mathcal{H}_2$, respectively, using \texttt{nthcomp}; $99.9\%$ and $73.9\%$ using \texttt{thcomp}; $99.98\%$ and $99.4\%$ using \texttt{highecut}; and $99.6\%$ and $64.4\%$ using \texttt{npex}.

\begin{table}
\centering
\begin{tabular}{c c c c c}
    Model               & $E_\text{cyc}$ [keV] & $\sigma$ [keV]  & $\tau$ & Significance           \\ \hline
    
    {$\mathcal{H}_1$}   & $52_{-5}^{+4}$       & $15_{-5}^{+3}$  & $10_{-7}^{+9}$        &  95.0$\pm$0.1 \%    \\ 
    \cline{2-5}   
    {$\mathcal{H}_2$}   & $48 \pm 4$            & $16_{-4}^{+3}$ & $13_{-7}^{+8}$        &       -     \\
                        & $19.5_{-0.9}^{+0.8}$  & $4 \pm 2$       & $0.3_{-0.2}^{+0.6}$   & 95.5$\pm$0.1 \%  \\ 
                        \hline
    
\end{tabular}
\caption{Best fit parameters and detection levels of the {\sc gabs} components derived from fitting {\it NuSTAR} spectra of 4U1700 with one ($\mathcal{H}_1$) and two ($\mathcal{H}_2$) components, with {\sc bbody+cutoffpl} model as the baseline continuum ($\mathcal{H}_0$).}
\label{cyctable}
\end{table}

Applying the $\textsc{cglumin}$ model of \textrm{XSPEC} in the 3--70~keV energy band, we calculated the source X-ray luminosity. Assuming a distance of ${\sim}1.9~\mathrm{kpc}$, we obtained an unabsorbed luminosity of $L_{\rm X} = (1.69\pm0.01){\times}10^{36}$~erg\,s$^{-1}$.

\subsubsection{Intensity-resolved spectroscopy}
We performed intensity-resolved spectroscopy on 4U1700 by segmenting the spectra according to the source count rate recorded in the full-spectrum light curves. Specifically, we divided the observation into three similar rate bands: a low-rate band (0--35~counts\,s$^{-1}$), a medium-rate band (35--55~counts\,s$^{-1}$), and a high-rate band (55--200~counts\,s$^{-1}$), with $\sim415-455\times10^3$ counts per band. For this analysis, we adopted the same absorbed blackbody plus cutoff power-law continuum model, including Zn and Cd edges, to fit the spectra over the 3--70~keV energy range. In the low-rate band, we measured a flux of $1.86_{-0.02}^{+0.01}{\times}10^{-9}$~erg\,cm$^{-2}$\,s$^{-1}$ with $\chi^{2}/\mathrm{dof}=1.06$; the medium-rate band yielded $5.73_{-0.06}^{+0.03}{\times}10^{-9}$~erg\,cm$^{-2}$\,s$^{-1}$ and $\chi^{2}/\mathrm{dof}=1.02$; and the high-rate band yielded $1.084\pm0.008{\times}10^{-8}$~erg\,cm$^{-2}$\,s$^{-1}$ and $\chi^{2}/\mathrm{dof}=1.09$. We examined the variability of the CRSFs across these rate-selected bands using \textsc{steppar} and $10^5$ spectral simulations, obtaining significances of $95.5\%$ for the soft band at $\sim$40~keV, $97.6\%$ for the medium band at $\sim$38~keV, and $98.8\%$ for the high band at $\sim$19~keV.


\section{Discussion} \label{sec:discussion}

The out-of-eclipse \textit{NuSTAR}\ observation of 4U1700 provides an excellent opportunity to study  the hard X-ray emission from this system. We do not detect coherent pulsation in the 100–5000~s range and constrain the pulsed fraction to $\mathrm{PF}_{90\%}\!<\!1.36$--$1.47\%$ (FPMA/FPMB; 3–79~keV), which disfavors the presence of stable, high-amplitude pulsations in this interval. The time-averaged spectrum is best described by a two-component continuum, \textsc{bbody+cutoffpl}, with an additional narrow Fe\,K$\alpha$ emission line component. On top of this baseline, we find model-dependent hints of absorption-like residuals at $\sim$19.5~keV and $\sim$52~keV that remain below the $3\sigma$ detection threshold.

\subsection{Comparison with previous X-ray campaigns}

Previous studies of the high-mass X-ray binary 4U1700 have reported strong flux variability on kilosecond timescales, commonly interpreted as the signature of clumpy wind accretion from the donor star \citep[e.g.,][]{Odaka2013, Garcia2018, Saavedra2022, Rikame2024MNRAS.529.3360R}. Our analysis confirms this behaviour, with flares ranging from a quiescent flux of ${\sim}2\times10^{-9}~\mathrm{erg\,cm^{-2}\,s^{-1}}$, consistent with \cite{vandermeer2005}, up to peaks of ${\sim}1\times10^{-8}~\mathrm{erg\,cm^{-2}\,s^{-1}}$, typical of sgHMXBs. As a zeroth-order estimate, one can relate the duration of individual flares to the characteristic size of the accreted wind clumps. Adopting an average flare duration of $\sim$1.5~ks (e.g. the interval between 36.8 and 38.4~ks in Fig.~\ref{4U1700_lc}) and using the binary parameters from \cite{Clark2002}, $M_{\rm star}{\sim}58$~M$_\odot$, $M_{\rm NS}{\sim}2.4$~M$_\odot$, and $P_{\rm b}{\approx}3.41$~d, Kepler's third law yields a mean orbital separation $a{\approx}2.6{\times}10^{7}$~km and a neutron-star orbital velocity $v_{\rm NS}{\approx}550$~km\,s$^{-1}$.

The wind velocity at the compact object position can be estimated from the \cite{1975CAK} velocity law, $v(r)=v_{\infty}(1-R_{\star}/r)^{\beta}$, where $r$ is the distance from the donor ($r=a$ for the NS), $\beta$ is in the range 0.5–1, and $v_{\infty} = 1750$~km\,s$^{-1}$ is the terminal wind velocity of HD~153919 \citep{Clark2002}. For $\beta=0.5$ we obtain $v_{\rm w}{\approx}1120$~km\,s$^{-1}$ at the neutron-star location. The relative velocity between the NS and the wind is then $(v_{\rm w}^2 + v_{\rm NS}^2)^{1/2}$, which, combined with the flare duration $\Delta t$, implies a characteristic clump size of $\Delta t\,(v_{\rm w}^2 + v_{\rm NS}^2)^{1/2} \approx 2{\times}10^{6}$~km, i.e. $\approx 0.07$~R$_\star$ \citep[$R_{ \star}\approx21.9$~R$_{\odot}$,][]{Clark2002}. This value is consistent with the clump sizes inferred from previous phenomenological studies of 4U1700 \citep{Garcia2019,Fogantini2021} and with hydrodynamical simulations of structured winds in sgHMXBs \citep{ElMellah2020}. No coherent pulsations are detected in the 0.01--5000~s range, in agreement with earlier non-detections based on EXOSAT and BATSE data, which placed upper limits of $0.5$--$5$\% \citep{Gottwald1986MNRAS.222P..21G} and ${\sim}4$\% \citep{Rubin1996}, respectively. Our upper limits on the pulsed fraction ($PF_{90\%}<1.5$--$1.8$\%) further constrain the presence of strong, stable pulsations in this interval.

In the spectral domain, several candidate CRSFs have been reported at ${\sim}16$–20~keV and ${\sim}38$~keV \citep{Reynolds1999,Bala2020MNRAS.493.3045B}, while a higher-energy feature around ${\sim}50$~keV has been suggested more recently \citep{Xiao2024}. These studies provide a useful reference for interpreting our out-of-eclipse \textit{NuSTAR} observations. When we fit the new \textit{NuSTAR} spectrum with the single absorbed \textsc{cutoffpl} continuum used in the earlier \textit{NuSTAR} analysis, the global fit is poor ($\chi^{2}/{\rm dof}=811.58/434$) and a clear excess remains at ${\sim}7.5$~keV that must be modelled as Fe\,K$\beta$. Adding a \textsc{gabs} component at ${\sim}$17--20~keV improves the statistic to $\chi^{2}/{\rm dof}=557.84/431$, corresponding to a $\sim\!98.2\%$ indication of a CRSF. 
However, replacing this single-component continuum with the two-component model that best represents our data (\textsc{bbody}$+$\textsc{cutoffpl}; see Section~\ref{sec:results}) removes systematic residual trends and yields a substantially better description of the spectrum. This demonstrates that the detailed residual pattern, rather than the global $\chi^{2}$ alone, must guide the choice of the null hypothesis $\mathcal{H}_{0}$ in CRSF searches.

The tentative CRSF at ${\sim}38$~keV reported by \citet{Reynolds1999} is not significantly detected in the new \textit{NuSTAR} epoch. Nonetheless, when we reanalyse the archival \textit{NuSTAR} dataset (ObsID~30101027002) with our preferred continuum and updated calibration, we recover an absorption-like feature at ${\sim}39$~keV with a detection probability of $\sim99.8\%$, improving $\chi^{2}/{\rm dof}$ from $446.64/417$ to $405.11/414$. In contrast, the ${\sim}16$~keV feature reported in the earlier \textit{NuSTAR} work only appears significant when a single-component continuum is imposed, underscoring the strong dependence of marginal features on the assumed continuum shape.
In the 2022 data, we find only weak ($\sim\!95\%$) evidence for an absorption feature at ${\sim}52$~keV (see Table~\ref{cyctable}), consistent with the higher-energy candidate proposed by \citet{Xiao2024}. The limited exposure and marked model dependence prevent us from claiming a firm detection of this or any other CRSF. 

It is also instructive to compare the line strengths. Previous works reported optical depths in the range of $\tau_{\rm cyc} \sim 0.3$--$0.6$ \citep{Reynolds1999, Bala2020MNRAS.493.3045B}. In our analysis, the marginal features observed at ${\sim}19.5$~keV and ${\sim}52$~keV correspond to central optical depths of $\tau_{\rm cyc} \approx 0.33$ and $\tau_{\rm cyc} \approx 0.27$, respectively (derived as $\tau \approx \text{Strength}/\sigma\sqrt{2\pi}$). These values align with the lower bound of historical claims but remain statistically insignificant in the context of the physically motivated \texttt{BBODY+CUTOFFPL} continuum, reinforcing the conclusion that line detection in 4U~1700-37 is critically dependent on the spectral state and continuum modeling.

Overall, the derived significances of candidate lines in 4U1700 can vary by more than $1\sigma$ depending on the adopted continuum and the dataset considered. This highlights the need for careful inspection of residuals, beyond global fit statistics, before asserting the presence of a cyclotron line, and indicates that deeper broadband observations with \textit{NuSTAR} or future hard X-ray missions will be required to reach the $>3\sigma$ level needed for an unambiguous confirmation of any CRSF in this system.

\subsection{Accretion regime in 4U1700}

The putative detection of CRSFs in X-ray spectra allows us to estimate the magnetic field strength of accreting NSs in X-ray binaries by applying of the standard formula of quantized Landau levels (see e.g., \citealp{Staubert2019}),

\begin{equation*}
\frac{B_{\rm cyc}} {10^{12}~{\rm G}} = \frac{1+z_g}{n}\left(\frac{E_\mathrm{cyc}}{\mathrm{11.57~keV}}\right),
\end{equation*}
where $0{\leqslant}z_g{\leqslant}0.3$ is the NS gravitational redshift \citep{Christodoulou2019} and $n$ the electrons excitation level.

Assuming $E_{cyc} = 19.5~\mathrm{keV}$ as the fundamental level ($n=1$) and no gravitational redshift ($z_{\rm g}=0$), we estimate a magnetic field strength of $B = 1.69_{-0.07}^{+0.08} \times 10^{12}~\mathrm{G}$. Alternatively, assuming $E_{cyc \,2} = 52~\mathrm{keV}$ with the same assumptions, we obtain a magnetic field strength of $B_2 = 4.4^{+0.3}_{-0.4} \times 10^{12}~\mathrm{G}$.

We note that the second CRSF component has a centroid energy that is more than two times (${\sim}$2.6) the energy of the first CRSF component. The detection of both components might be attributable to the scattering by electrons located in two significantly distant regions within the accretion column.This deviation from a strict harmonic integer ratio is reminiscent of other wind-fed systems like Vela X-1, which exhibits complex line profiles or anharmonic features often interpreted as the superposition of resonant scattering at different altitudes and optical depths within the column \citep{Staubert2019}.

The measured X-ray luminosity of 4U1700 allows us to infer the rotation period of the NS by applying the accretion model developed by \cite{Shakura2012} for slowly rotating magnetized NS with X-ray luminosities below ${\sim}4{\times}10^{36}~\mathrm{erg\,s^{-1}}$. In this model of quasi-spherical subsonic accretion, the accreted matter subsonically settles down onto the rotating magnetosphere forming an extended quasi-static shell. This shell mediates the angular momentum removal from the rotating NS magnetosphere during spin-down episodes by large scale convective motions. The accretion rate through the shell is determined by the ability of the plasma to enter the magnetosphere.

The settling regime of accretion which allows angular momentum removal from the NS magnetosphere can be realized for moderate accretion rates $\dot M _{16} < 3.7\mu_{30}^{1/4}$. In this expression, $\dot M_{16} \equiv \dot M/10^{16}~\mathrm{g\,s^{-1}}$ is the normalized mass accretion rate and $\mu_{30} \equiv \mu/10^{30}~\mathrm{G\,cm^{3}}$ is the normalized magnetic dipole moment of the X-ray pulsar related to the surface equatorial dipole magnetic field. At higher accretion rates, a free-fall gap above the NS magnetosphere appears due to rapid Compton cooling, and accretion becomes highly non-stationary.

Taking the previously derived value for the magnetic field and considering a typical NS radius of 10~km, we find that $\mu_{30}=0.5 B R_{\rm NS}^{3}= 0.84_{-0.04}^{+0.03}$. Thus, the critical mass accretion rate for 4U1700 yields $\dot M _{16}{\simeq}3.5$. The observed X-ray luminosity of 4U1700 previously obtained translates to a normalized accretion rate of approximately $\dot M _{16}\sim10\,L_{X}/c^{2} = 1.88\pm0.01$, where $c$ is the speed of light. We see then that 4U1700's estimated accretion rate is below the supersonic Bondi-Hoyle-Littleton accretion rate, thus satisfying the conditions for applying the framework of quasi-spherical subsonic accretion.

The equilibrium period of the pulsar may be computed as
\begin{equation*}
    P_{\text{eq}}\simeq 940\mathrm{[s]} \, \mu_{30}^{12/11} \left(\frac{P_b}{10\text{d}}\right){\dot M_{16}^{-4/11}} v_{8}^{4},
\end{equation*}
where $P_{\text{b}}\simeq 3.411581~\mathrm{d}$, and $v_{8} \equiv v_{w}/10^{8}\mathrm{[cm\,s^{-1}]}$ where $v_{w}$ is the stellar wind velocity. Considering $v_{w}=1750~\mathrm{km\,s^{-1}}$ for the stellar companion HD 153919 \citep{Clark2002}, we obtain an equilibrium period of $P_{\text{eq}}\simeq 1.9 \pm 0.1~{\mathrm{ks}}$.

Performing a similar analysis using the $\sim$39~keV CRSF detected in the archival observation, which we consider as the harmonic (n=2, corresponding to the second Landau level) of the $\sim$19.5~keV CRSF recently detected, and adopting a luminosity of $L_{\rm X,2} = (8.16\pm0.05){\times}10^{35}$~erg\,s$^{-1}$, we obtain $P_{\text{eq,2}}\simeq 2.5_{-0.1}^{+0.2}~{\mathrm{ks}}$, where the uncertainties correspond to a $1\sigma$ significance level.

Expanding the uncertainties to 4$\sigma$, we obtain that the two calculations range from 2.2--3.0~ks and 1.5--2.3~ks, respectively, coinciding in the 2.2--2.3~ks range. This difference might be due to the calculations being performed for different orbital phases, and thus different X-ray luminosities between the two observations. We note that while 4U1700 is variable, these estimates are robust: since $P_{eq} \propto \dot{M}^{-4/11}$, even an order of magnitude variation in the long-term average luminosity would only shift $P_{eq}$  by a factor of $\sim$2.3 (as seen between our two values here), keeping it in the kilosecond regime.

As presented in Section~\ref{sec:results}, we do not detect any coherent pulsation in the \textit{NuSTAR}\ light curves of 4U1700 above a $3\sigma$ significance level. This could result from the geometric configuration of the rotational and dipole moment axes with respect to the line of sight. In the case of a complete non-crossing with the line of sight, our spin period estimation (derived from hints of CRSFs on {\it NuSTAR} spectra) would place 4U1700 in the upper-left region of the Corbet diagram \citep{Corbet1986}, within the sgHMXB cluster, consistent with its known classification derived from the spectral properties of its stellar donor. On the other hand, \citet{Xiao2024} suggested that the pulsations could be intermittent, manifesting only within brief intervals, as observed in LS~+61$^\circ$303 \citep{Weng2022}.


\section{Conclusions} \label{sec:conclusions}

Our findings from the analysis of a $\sim$50~ks \textit{NuSTAR}\ observation of 4U1700 can be summarized as follows:

\begin{itemize}
    \item No coherent pulsations were detected in the \textit{NuSTAR} light curves with significances above $3\sigma$ in the 0.01–5000~s range. We place $90\%$ upper limits on the pulsed fraction of $1.36\%$ and $1.47\%$ for FPMA and FPMB, respectively, in the 3–79~keV band.
    \item A systematic search for CRSFs in both available \textit{NuSTAR} epochs, combining several continua and extensive spectral simulations, reveals absorption-like features at ${\sim}19.5$, ${\sim}39$ and ${\sim}52$~keV, whose presence and centroid energy are strongly epoch- and model-dependent. Under our best-fitting continuum, none of these features reaches the conventional ${\geq}3\sigma$ detection threshold, but their detection probabilities can exceed $99.9\%$ in alternative continua with formally worse global fits.
    \item Interpreted as cyclotron features, the ensemble of candidate lines and non-detections constrains the neutron star magnetic field to $B\sim(1.7$–$4.4)\times10^{12}$~G, disfavouring both very low fields and magnetar-like values. Within the quasi-spherical settling accretion framework, the measured luminosity and this magnetic-field range imply an equilibrium spin period $P_{\rm eq}\approx1.9$~ks, placing 4U1700 in the supergiant HMXB region of the Corbet diagram and consistent with the lack of detected pulsations.
    \item Our analysis illustrates that the significance and centroid energy of candidate CRSFs in wind-fed HMXBs can be strongly dependent on the adopted continuum, calibration and flux state. Multi-epoch, multi-continuum analyses combined with simulation-based significance estimates are therefore essential to extract robust physical information from current and future hard X-ray observations of systems like 4U1700.
\end{itemize}


\section*{Acknowledgements}
Funded by the European Union (Project 101183150 - OCEANS). Views and opinions expressed are however those of the author(s) only and do not necessarily reflect those of the European Union or the European Research Executive Agency (REA). Neither the European Union nor REA can be held responsible for them. FAF, JAC, and FG acknowledge support by PIP 0113 (CONICET). EAS acknowledges support by the Spanish \textit{Agencia estatal de investigaci\'on} via PID2021-124879NB-I00 and PID2024-161863NB-I00. LWO is an undergraduate student at FCAG and a fellow of the Consejo Interuniversitario Nacional, Argentina. FAF is a postdoc fellow of CONICET. JAC, FG and JFAC are CONICET researchers. JAC, PLLE and JM were supported by Consejería de Economía, Innovación, Ciencia y Empleo of Junta de Andalucía as research group FQM-322. JAC, FG, JM and PLLE were also supported by grant PID2022-136828NB-C42 funded by the Spanish MCIN/AEI/ 10.13039/501100011033 and “ERDF A way of making Europe”.


\end{document}